\documentclass{aa}  

\usepackage{graphicx}
\usepackage[varg]{txfonts}
\usepackage{natbib}
\usepackage{float,caption,afterpage}
\usepackage{natbib,twoopt}
\usepackage[breaklinks=true]{hyperref} %% to avoid \citeads line fills
\bibpunct{(}{)}{;}{a}{}{,} %% natbib format for A&A and ApJ
\makeatletter
\newcommandtwoopt{\citeads}[3][][]{\href{http://adsabs.harvard.edu/abs/#3}%
{\def\hyper@linkstart##1##2{}%
\let\hyper@linkend\@empty\citealp[#1][#2]{#3}}}
\newcommandtwoopt{\citepads}[3][][]{\href{http://adsabs.harvard.edu/abs/#3}%
{\def\hyper@linkstart##1##2{}%
\let\hyper@linkend\@empty\citep[#1][#2]{#3}}}
\newcommandtwoopt{\citetads}[3][][]{\href{http://adsabs.harvard.edu/abs/#3}%
{\def\hyper@linkstart##1##2{}%
\let\hyper@linkend\@empty\citet[#1][#2]{#3}}}
\newcommandtwoopt{\citeyearads}[3][][]%
{\href{http://adsabs.harvard.edu/abs/#3}
{\def\hyper@linkstart##1##2{}%
\let\hyper@linkend\@empty\citeyear[#1][#2]{#3}}}
\makeatother

\begin{document}

   \title{The LBV HR Car has a partner\thanks{Based on data obtained with ESO programmes 092.C-0243, 092.D-0289,  092.D-0296,  094.D-0069, and 596.D-0335.}}

 \subtitle{Discovery of a companion with the VLTI}

   \author{Henri M.J. Boffin\inst{1,2}         
                       \and 
                       Thomas Rivinius\inst{1} 
                       \and
                       Antoine M\'erand\inst{1}  \and
       Andrea Mehner\inst{1}
\and
           Jean-Baptiste LeBouquin\inst{3}
           \and
           	Dimitri Pourbaix\inst{4 \fnmsep\thanks{Senior Research Associate, F.R.S.-FNRS, Belgium}}
	\and
       Willem-Jan de Wit\inst{1}
             \and
       Christophe Martayan\inst{1}
       \and
             Sylvain Guieu\inst{1,3}
                  }

   \institute{ 
 ESO, Alonso de C\'ordova 3107, Casilla 19001, Santiago, Chile\\
              \email{hboffin@eso.org}
              \and
              ESO, Karl-Schwarzschild-str. 2, 85748 Garching, Germany
         \and
             Institut de Plan\'etologie et d'Astrophysique de Grenoble (UMR 5274),  BP 53, 38041 Grenoble C\'edex 9, France 
             \and
             	Institut d'Astronomie et d'Astrophysique, Universit\'e Libre de Bruxelles (ULB), Belgium
             }

  % \date{Received March 15, 2014; accepted March 16, 2014}

\authorrunning{H.M.J. Boffin et al.}
 %\abstract{}{}{}{}{} 
% 5 {} token are mandatory
\abstract{
Luminous Blue Variables (LBVs) are massive stars caught in a post-main sequence phase, during which they are losing a significant amount of mass. 
As, on one hand, it is thought that the majority of massive stars are close binaries that will interact during their lifetime, and on the other, the most dramatic example of an LBV, $\eta$~Car, is a binary, it would be useful to find other binary LBVs.
We present here interferometric observations of the LBV HR\,Car done with the AMBER and PIONIER instruments attached to ESO's Very Large Telescope Interferometer (VLTI). Our observations, spanning two years, clearly reveal that HR\,Car is a binary star. It is not yet possible to constrain fully the orbit, and the orbital period may lie between a few years and several hundred years. We derive a radius for the primary in the system and possibly resolve as well the companion. The luminosity ratio in the $H-$band between the two components is changing with time, going from about 6 to 9.  We also tentatively detect the presence of some background flux which remained at the 2\% level until January 2016, but then increased to 6\% in April 2016. 
Our AMBER results show that the emission line forming region of Br$\gamma$ is more extended than the continuum emitting region as seen by PIONIER and may indicate some wind-wind interaction.
Most importantly, we constrain the total masses of both components, with the most likely range being 33.6~M$_\odot$ and 45~M$_\odot$. Our results show that the LBV HR\, Car is possibly an $\eta$~Car analog binary system with smaller masses, with variable components, and further monitoring of this object is definitively called for. 
}


   \maketitle
%
%________________________________________________________________

\section{Introduction}
Luminous Blue Variables (LBVs) are post-main-sequence massive stars undergoing a brief, but essential, phase in
  their life, characterised by extreme mass-loss and strong photometric and spectroscopic variability \citepads{1984IAUS..105..233C,1994PASP..106.1025H,2012ASSL..384..221V}. The best known LBV, \object{$\eta$\,Carinae}, is
  known to have lost about 10--30 solar masses during its great outburst in the 1840s \citepads{2003AJ....125.1458S}. There is, as yet, no firmly established
  mechanism to explain the large mass
  loss of LBVs, nor how it happens: is  
 the mass lost due to a steady radiatively-driven stellar wind, or is it removed by punctuated eruption-driven mass
loss, such as the great outburst of $\eta$\,Car? Numerous hypotheses have been
  proposed. Apart from single star processes, such as core and atmospheric instabilities \citepads{1994PASP..106.1025H} or supercritical rotation,
 the binary hypothesis \citepads[e.g.][]{1989ASSL..157..185G,2011MNRAS.415.2020S} is a very strong contender, especially
 as it is well established that massive stars form nearly exclusively in multiple systems and that binary interactions
 are critical for these stars \citepads{Chini2012,Sana2012,Sana2014}.  
 
There is currently a hot debate in the literature on the evolutionary status of LBV stars and on the importance of binarity in their formation \citepads{2015MNRAS.447..598S,2016arXiv160301278H}. 
So far, however, while several wide LBV binaries were identified, LBV systems similar to $\eta$\, Car (relatively close \& eccentric) have not been found \citepads{Mar2012,2016A&A...587A.115M}.  The only possible exception might be the LBV candidate \object{MWC 314} \citepads{Lobel2013}, but this is apparently a massive semi-detached binary system, and thus not directly comparable to $\eta$\,Car. On the one hand, this may appear rather surprising as it is thought that given their very high multiplicity rate, more than 70\% of all massive stars will exchange mass with a companion \citepads{Sana2012}. On the other hand, LBVs are rare objects with complex emission line spectra and intricate nebulae. Located at average distances of a few kpc or more, they therefore require at least milli-arcsecond resolution for direct close companion detection. Such a resolution is only reachable by interferometry.
 
 %\tikzsetnextfilename{fig_HRCar_AMBER}
\begin{figure*}[htbp]
\begin{center}
\includegraphics[width=16cm]{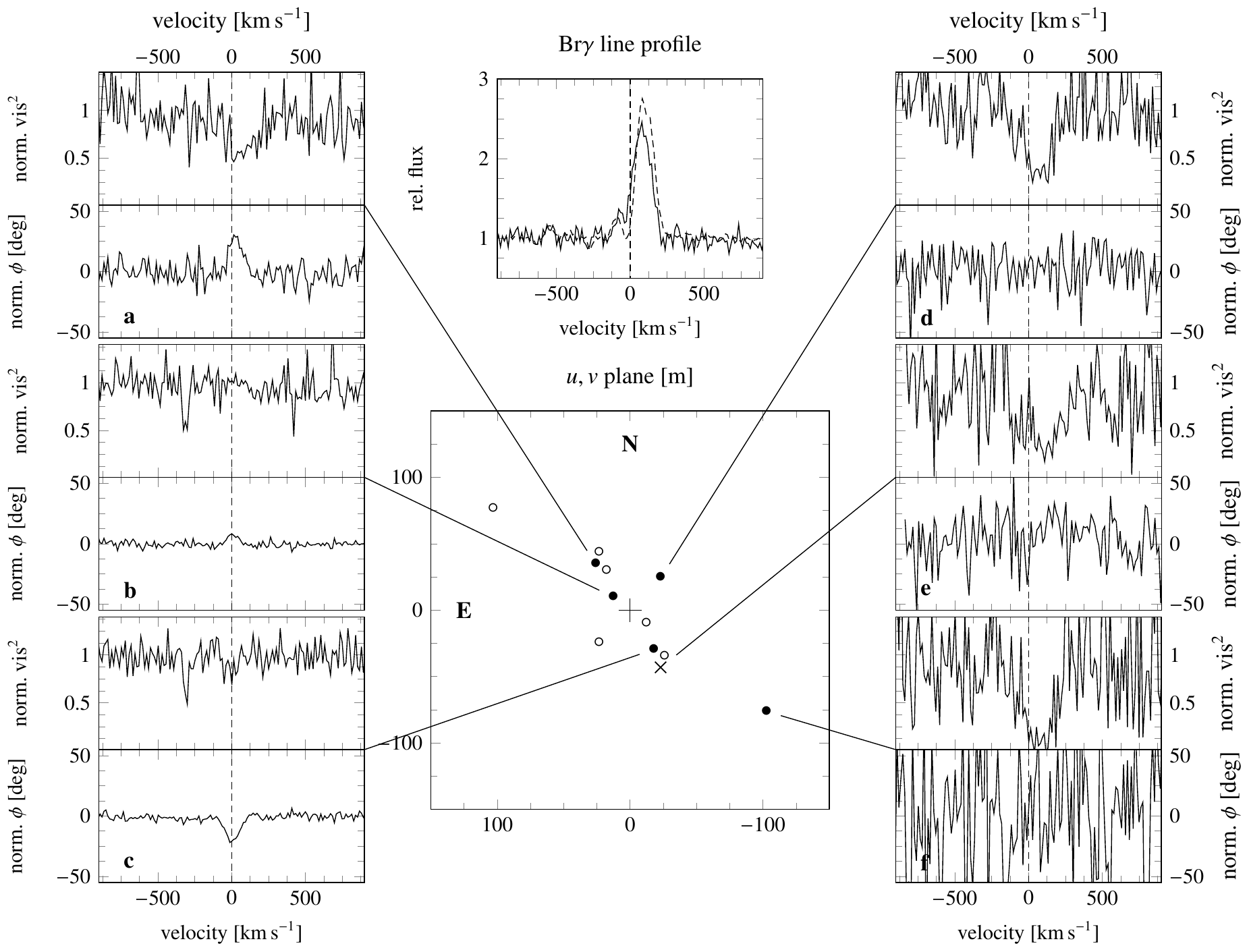}
\end{center}
\caption[xx]{\label{fig:AMBER}Exemplary AMBER observations of HR\,Car. 
In the $u,v$-plane coverage, shown in the center, solid lines connect
visibility and phase observation with their respective $u,v$ points, open
circles are conjugated $u,v$ points. Panels a) and d) show observations obtained
on MJD\,56718.149, panels b) and c) on MJD\,56676.226, and panel f) on MJD\,
56726.113. One baseline from the observation taken on MJD\,55311.024 is shown
as dashed line in the flux panel and in panel e), its $u,v$ position is marked
by a cross. }
\end{figure*}	

Here, we report on interferometric measurements of the LBV \object{HR\,Carinae} (HD\,90177), one of the very few in the Milky Way.  \citetads{1990A&AS...82..189V} derived for this star an effective temperature of 14\,000$\pm$2\,000 K and a bolometric luminosity (M$_{bol}$) of $-9.5$, with a mass-loss rate of 2.2~10$^{-5}$~M$_\odot$yr$^{-1}$. The luminosity was revised to M$_{bol}=-8.9$ and the distance to 5$\pm$1 kpc by \citetads{1991A&A...246..407V}, putting HR\,Car most likely in the Carina spiral arm. At the same time, \citetads{1991A&A...248..141H} derived a kinematic distance of  5.4$\pm$0.4~kpc and showed that HR\,Car has a multiple shell expanding atmosphere. \citetads{1997A&A...320..568W} found that HR\,Car has a nebula that appears bipolar, with each lobe having a diameter of $\sim$0.65~pc and a line-of-sight expansion velocity of 75--150 kms$^{-1}$. We note in passing that the Hipparcos measurement of the parallax of HR\,Car of 
 1.69$\pm$0.82 mas  \citepads{2007A&A...474..653V}, translating to a very imprecise distance of 592$^{+557}_{-193}$ pc, is most likely incorrect in views of the other indicators, and quite possibly a result of the hitherto unknown binarity.

Effective temperature determinations for HR\,Car  range between about 10\,000 K \citepads{2002A&A...387..151M},  14\,000$\pm$ 2\,000 K (see above), 17\,900 K \citepads{2009ApJ...705L..25G}, and 22\,000 K \citepads{2010AN....331..349H}. 
The star is highly variable: it had its last S~Dor outburst in July 2001 and is
currently in a quiet state, two magnitudes fainter than at maximum (in $V$). Visual magnitudes obtained on the AAVSO web site indicate indeed that the star has now a magnitude $V\sim 8.7-9$.  
According to \citetads{2011MNRAS.410..190T}, HR\,Car is a B2evar star with a 
mass of 18.1$\pm$5.5~M$_\odot$ and an age of 5.0$\pm$1.4 Myr, while 
\citetads{2010AN....331..349H}  quote a value of 23.66$\pm$7.24~M$_\odot$. Similarly, \citetads{2009ApJ...705L..25G} show the star to have a high rotational velocity of $150\pm20$ kms$^{-1}$, i.e. rotating at 88\% of its critical velocity, and derive a current mass of about 25~M$_\odot$, and an initial mass of 50$\pm$10~M$_\odot$, but we should stress here that all these values are very model-dependent. The high velocity and the difficulty of LBVs to lose angular momentum led \citetads{2009ApJ...705L..25G} to suggest that HR\,Car could explode during its current LBV phase, making the link with detections of LBV-like progenitors of Type IIn supernovae \citepads{2015MNRAS.447..598S}. It is thus important to characterise as best as possible this very interesting star.

We show here, based on interferometric measurements, that the LBV HR\,Car is in fact a binary system, with an orbital period of several years, making it therefore the first LBV similar to $\eta$~Car in terms of binarity. Our observations are presented in Sect.\ref{Sec:Obs} and discussed in Sect.\ref{Sec:Dis}.

\section{Observations}\label{Sec:Obs}
\subsection{AMBER}
The LBV HR Car was observed as part of the OHANA survey
\citepads{2016arXiv160203457R}, which secured spectrally resolved 3-beam
interferometry of Br$\gamma$ with AMBER \citepads{2007A&A...464....1P} at ESO's Very Large Telescope
Interferometer (VLTI).
In addition, a single previous AMBER observation of HR\,Car was obtained from
the archive.  All AMBER observations were reduced the standard way, i.e.\ with
{\tt amdlib}\footnote{\url{http://www.jmmc.fr/data_processing_amber}}, in version 3.0.6 \citepads{2009A&A...502..705C}. A summary of the
observations is given in Table~\ref{tab:AMBER}.  The way the OHANA survey was
designed, no dedicated calibrators were taken, and hence the visibility and
phase observations are normalised to the local continuum.  Due to the backup
and snapshot nature of the OHANA survey, the data quality is not very homogeneous, as
can be seen in Fig.~\ref{fig:AMBER}, where some of the AMBER observations are
shown.

The observed Br$\gamma$ line profile is typical for strong LBV winds, with a
P\,Cyg type absorption seen at about zero and negative velocities, though not
descending below the level of the continuum, and the peak of the emission
somewhat redshifted.

The visibility signature, in first order a measure of the size of the emission
region, is well centred on the overall emission. At the shortest baselines,
shown in panel c), the visibility drop is too small to be seen, but along the
NE-SW axis it becomes noticeable at about 30 to 40\,m baseline length, and the
visibility has reached zero at above 100\,m baseline length. In the $K$-band,
this means the line emission region of Br$\gamma$ is of the order of 5\,mas in size in
this direction.  Only few measurements were aligned in the NW-SE direction,
but  in panel d) of Fig.~\ref{fig:AMBER} it can be seen that the drop in
visibility is stronger than on similar baseline lengths in the perpendicular
direction, shown in panels a) and c). This means that the  Br$\gamma$ wind emission is not
spherical in size, and more extended along the NW-SE axis.

\begin{figure}[htbp]
\includegraphics[width=9cm]{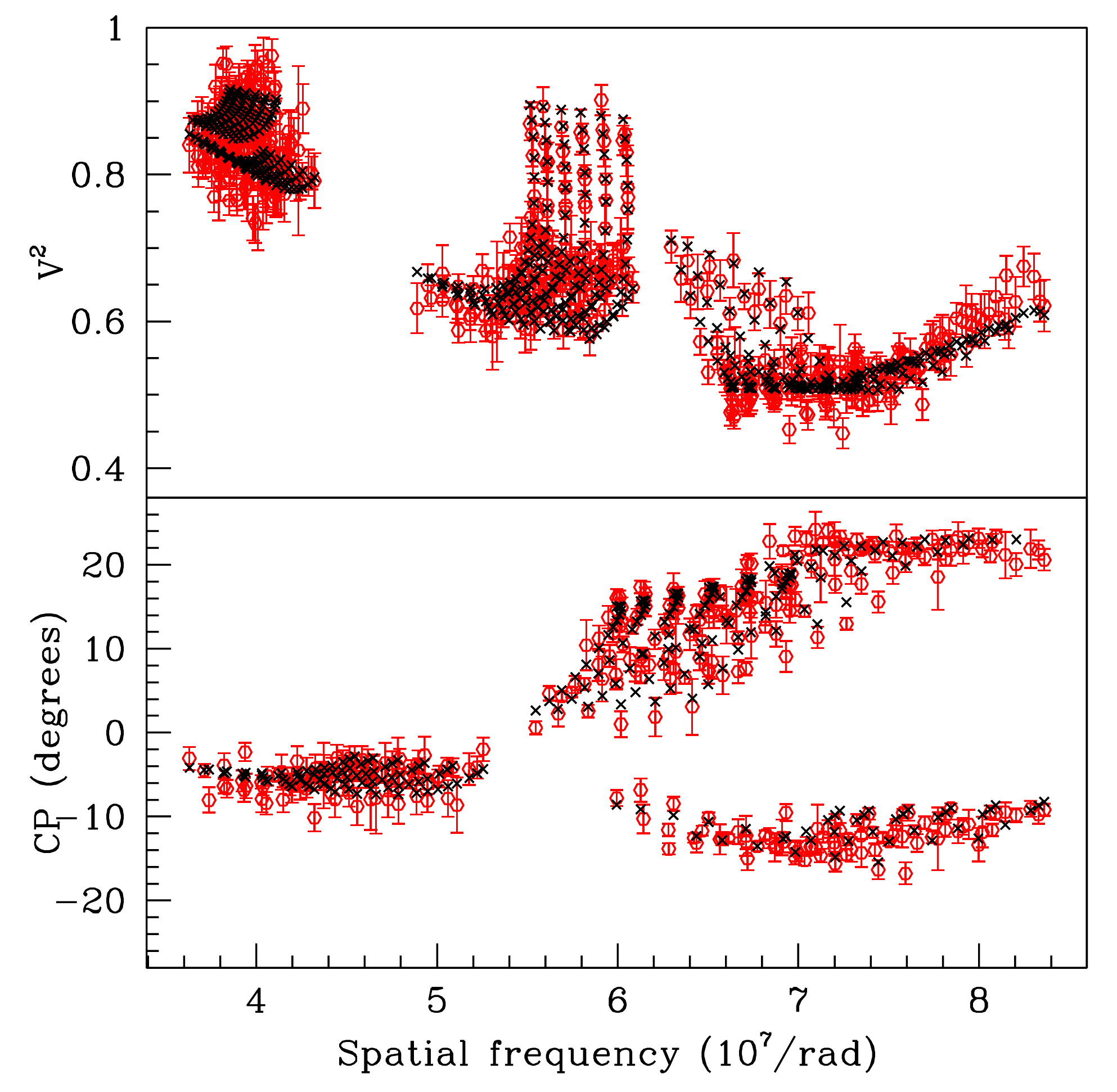}
\caption{\label{Fig:vis2pionier} PIONIER observations of HR\,Car in January 2015: squared visibilities (top) and closure phases (bottom). The data are shown as red crosses with error bars, while the best fit model of two discs on top of a background 
is shown with black crosses. A single star would have bijective visibilities and a 0$\pm$180$\deg$ closure phase.
}
\end{figure}

In
first order approximation, wavelength-differential phase, the second interferometric observable, is a measure of  photocentre shift. A  differential phase signature in the emission line is observed along the
NE-SW axis, but not along the NW-SE axis, meaning in the NW-SE direction the
Br$\gamma$ emission is symmetric to the center-of-light in the continuum. In
the NE-SW direction, however, it is offset towards the NE. Other than the
visibility profile, the phase signature is also {\em not} centered on the
emission in wavelength, but shifted to the blue. In other words, the part of
the emission that is offset to the NE is as well approaching the observer.

The signatures of phase and visbility observed in the OHANA survey are
temporally stable, i.e.\ over the several months of observations. In
particular the phase signature, hallmarking a photocenter displacement, is
hard to explain with the variable, yet more-or-less symmetric wind of a single
supergiant star. 

On longer timescales, however, there are changes. One of the three baselines
observed in 2010 was at a similar $u,v$ position as some of the ones observed
in 2014, see Fig.~\ref{fig:AMBER} panels a) and e) vs.\ c). Even though the
quality of the data was lower in 2010, there was a clear visibility drop at
this $u,v$ position, but little to no phase signature in 2010, while in 2014
this was quite the opposite.

The above description is hard to reconcile with a single star model, and
prompted further observations with another interferometric instrument,
PIONIER, to test a binary hypothesis \citepads[see][for a discussion of the
  alternatives and initial justification of the binary
  hypothesis]{2015IAUS..307..295R}.

\begin{table}
\caption{\label{tab:AMBER}Log of AMBER observations.}
\begin{center}
\begin{tabular}{lll}
\hline
\hline
Date & Stations  & MJD         \\
\hline
April 25, 2010 & U1--U2--U4& 55311.024  \\
January 10, 2014 & A1--B2--C1& 56667.361  \\
January 19, 2014 & A1--C1--D0& 56676.226  \\
February 8, 2014 & D0--H0--G1& 56696.343  \\
February 9, 2014 & A1--B2--C1& 56697.284  \\
February 12, 2014 & A1--B2--C1& 56700.172  \\
February 22, 2014& D0--H0--I1& 56710.103  \\
March 2, 2014 & D0--H0--I1& 56718.085  \\
March 2, 2014& G1--I1--H0& 56718.149  \\
March 10, 2014 & A1--G1--K0& 56726.113  \\
\hline
\end{tabular}
\end{center}
\end{table}

% Requires the booktabs if the memoir class is not being used
\begin{table}[htbp]
    \caption{Log of the PIONIER observations. We indicate the time of observation, the number of visibilities and closure phases obtained, as well as the mean value of the epoch of observation.}
   \label{tab:log}
    \centering
   \begin{tabular}{@{} lccl @{}} % Column formatting, @{} suppresses leading/trailing space
      \hline\hline
      Date    & $V^2$ points & CP points& MJD \\
      \hline
      March 2, 2014 & 53 & 35 & 56718.254\\
       March 3, 2014 & 18 & 12 & 56719.047\\
      January 26, 2015 & 252 & 168 & 57048.310 \\
      January 5, 2016 & 102 & 68 & 57392.337\\
      February 1, 2016 & 90 & 60 & 57419.118\\
      April 3, 2016 & 71 & 48 & 57482.060\\
      \hline
   \end{tabular}
\end{table}

\begin{table*}[th]
   \centering
    \caption{  \label{tab:models} Parameters of the best {\tt LITpro} models to the PIONIER data. The table gives the angular diameter of the primary, and for the secondary, its relative flux fraction, angular diameter, and its position, as well as the background flux if present. All dimensions are given in mas.}
   \begin{tabular}{@{} lcccccccr @{}} % Column formatting, @{} suppresses leading/trailing space
        \hline\hline
\\
%\multicolumn{8}{c}{LITPro} \\
Date & Model type &  {Primary} & \multicolumn{4}{c}{Secondary} & Background & $\chi^2_r$ \\
 & & Diameter & Flux fraction & Diameter & $x$ & $y$ & Flux fraction &\\
\hline
\\
{Mar 2014} 
& 2 discs + bcg & 0.37$\pm$0.20 &  14$\pm$1.5\%&1.03$\pm$0.41  &  $-1.08\pm$0.03 &$-1.70\pm$0.03  & 1.2$\pm$0.4\% &1.0\\
\\
{Jan 2015} & 
2 points &  --  & 14.0$\pm$0.6\% & --  & $-1.54\pm$0.01 & $-0.71\pm$0.01 & -- & 3.6\\
& 2 discs &  0.59$\pm$0.03  & 13.8$\pm$0.7\% & 1.23$\pm$0.09 & $-1.55\pm$0.01 &$-0.76\pm$0.01  & -- &  1.8\\
& 2 points + bcg & -- & 12.6$\pm$0.6\% & -- & -1.52$\pm$0.01 & $-0.70\pm$0.01 & 3.2$\pm$0.2\% &1.6\\
& 1 point + disc + bcg & -- & 14$\pm$0.6\% & 1.07$\pm$0.09 & -1.50$\pm$0.01 & $-0.69\pm$0.01 & 2.5$\pm$0.2\% &1.55\\
& 2 discs + bcg & 0.39$\pm$0.04 & 12.9$\pm$0.6\% & 0.76$\pm$0.15 & -1.55$\pm$0.01 & $-0.73\pm$0.01 & 2.2$\pm$0.2\% &1.5\\
\\
{Jan 2016} 
& 2 discs + bcg &  0.45$\pm$0.09&13$\pm$3\%&0.86$\pm$0.22&$-0.68\pm0.07$&0.89$\pm$0.09& 1.6$\pm$0.4\% & 0.8\\
\\
{Feb 2016} 
& 2 discs + bcg &  0.30$\pm$0.09&16$\pm$2\%&0.86$\pm$0.22&$-0.53\pm0.02$&0.91$\pm$0.03& 3.5$\pm$0.4\% & 0.8\\
\\
{Apr 2016} 
& 2 discs + bcg &  0.37$\pm$0.04&9.0$\pm$0.6\%&0.87$\pm$0.11&$-0.39\pm0.01$&1.35$\pm$0.03& 4.7$\pm$0.3\% &0.6\\
\\\hline
\end{tabular}
\end{table*}

\subsection{PIONIER}
We observed HR\,Car with the four 1.8-metre Auxiliary Telescopes of the VLTI, using the PIONIER visitor instrument \citepads{berger_2010,2011A&A...535A..67L} in the $H$-band on the nights of 1--2, and 2--3 March 2014. This provided us with strong hints of the binarity of HR\,Car, and we therefore started to observe it on a regular basis. The log of our observations is shown in Table~\ref{tab:log}. 
In 2014, we used the prism in low resolution (SMALL), the fringes being sampled over three spectral channels and the intermediate VLTI configuration D0-H0-I1-G1 was used. In 2015, we used the extended configuration A1-G1-K0-I1 and in 2016, the A0-G1-J2-J3 one. Early 2015, the detector of PIONIER was upgraded and from then on we used the GRISM mode, where the fringes are sampled over six spectral channels. The stars HD~90074, HD~87238 and HD~90980 were used as calibrators, in successive CAL-SCI-CAL sequences, with five SCI images taken during each sequence. As we are using four telescopes, we have in general for each observations six visibilities and four closure phases (times the spectral channels). 
As we had only a few data points on the night of 2--3 March 2014, we have combined them with those obtained the night before.

\begin{figure*}[htbp]
\begin{centering}
\begin{tabular}{cc}
\includegraphics[width=8.5cm]{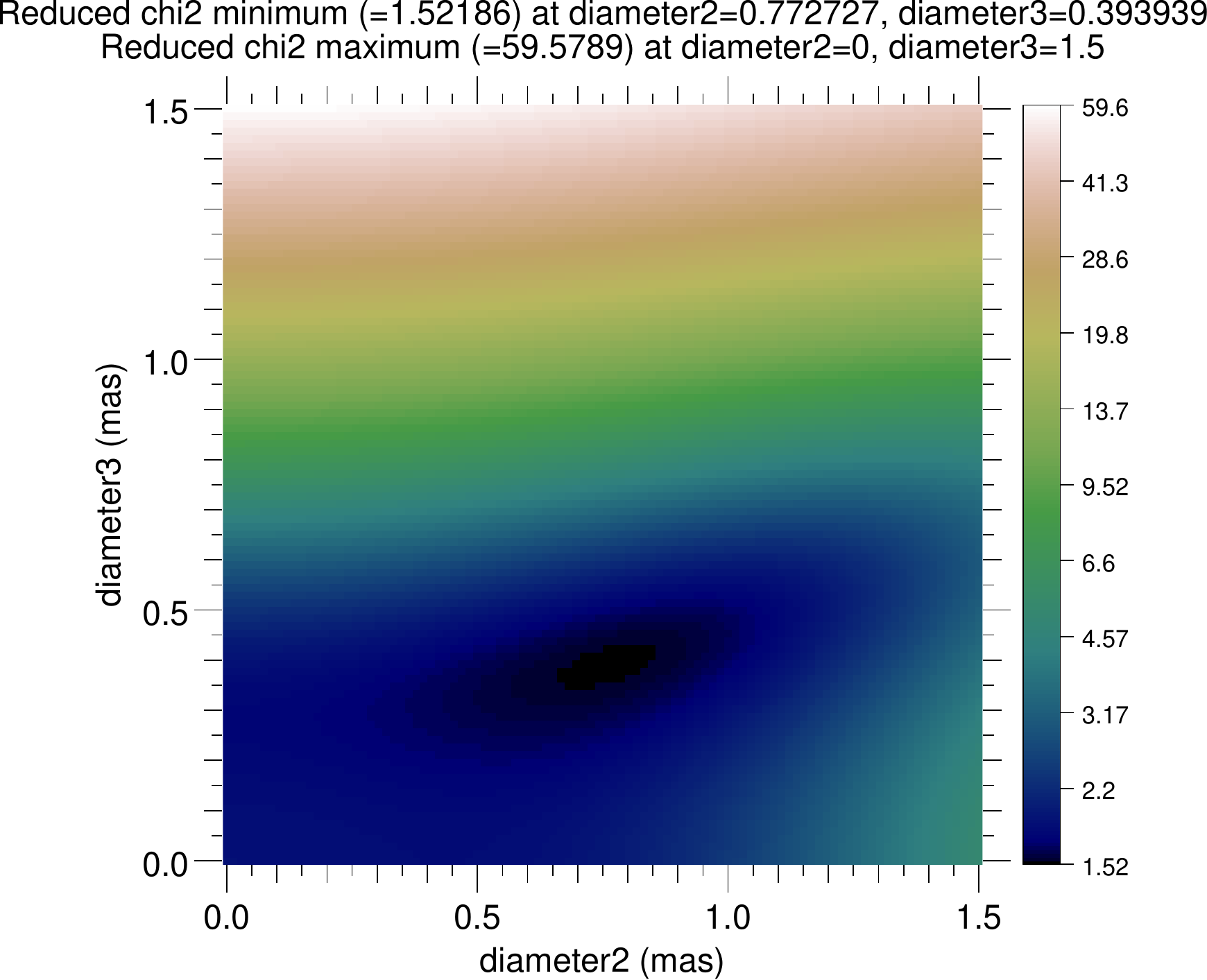} &
\includegraphics[width=8.5cm]{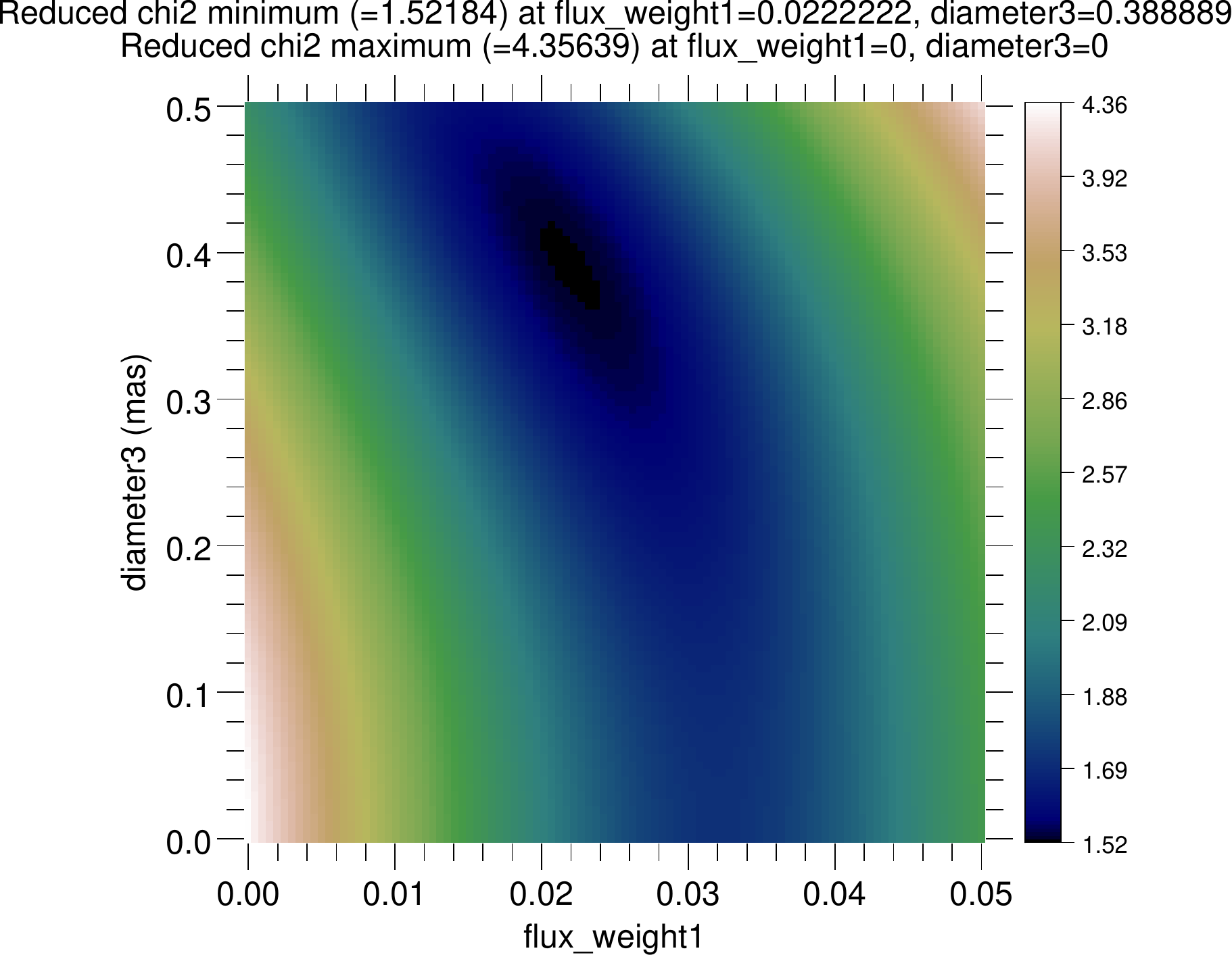}\\
\end{tabular}
\caption{\label{Fig:chi2map} (right) Reduced $\chi^2$-map in the diameter-diameter plane for the two-disc+background {\tt LITpro} model for the January 2015 data. (left) Same but in the primary disc diameter-background flux plane. The maps show the confidence interval for the best fit parameters.
}
\end{centering}
\end{figure*}

Data reduction was done in the usual way with the {\tt pndrs}\footnote{\url{http://www.jmmc.fr/data_processing_pionier.htm}} package presented by Le Bouquin et al. (2011). We show the resulting squared visibilities and closure phases for the best data set -- the one from January 2015 -- in Fig.~\ref{Fig:vis2pionier}.

\begin{table*}[th]
   \centering
    \caption{  \label{tab:candid} Parameters of the best {\tt CANDID} models to the PIONIER data. The table gives the angular diameter of the primary, the background flux, and for secondary, its relative flux fraction and its position. The last column is the reduced $\chi^2$ of the model. All dimensions are given in mas. }
   \begin{tabular}{@{} llccccr @{}} % Column formatting, @{} suppresses leading/trailing space
        \hline\hline
\\
Date &     {Primary} & Background & \multicolumn{3}{c}{Secondary} & $\chi^2_r$ \\
& Diameter & Flux fraction & Flux fraction & $x$ & $y$ & \\
\hline
\\
{ Mar 2014} & 
 $0.38^a$ & $1.7^{+1.2}_{-0.9}$\%& $16.4^{+1.0}_{-0.8}$\%& $ -1.06^{+0.04}_{-0.04} $&$-1.70^{+0.07}_{-0.07}$& 2.30\\
\\
{ Jan 2015} & 
$0.45^{+0.07}_{-0.05}$& $2.7^{+0.6}_{-0.6}$\%& $14.3^{+0.3}_{-0.4}\%$ & $-1.57^{+0.01}_{-0.01}$ & $-0.74^{+0.03}_{-0.02}$& 1.53 \\
\\
{ Jan 2016} & 
$0.34^{+0.17}_{-0.12} $& $2.2^{+0.7}_{-0.6}$\% &$12.2^{+3.8}_{-5.7}$\%&$  -0.79^{+0.14}_{-0.11}$& $ 0.93^{+0.13}_{-0.22}$& 1.45 \\
\\
{ Feb 2016} &
$0.37^{+0.13}_{-0.11} $& $4.3^{+0.7}_{-0.6}$\% &$14.1^{+1.3}_{-1.9}$\%&$-0.58^{+0.05}_{-0.05}$& $ 1.02^{+0.05}_{-0.04}$& 1.30\\
\\
{ Apr 2016} &
 $0.37^{+0.07}_{-0.05} $& $5.8^{+0.6}_{-0.5}$\% &$9.9^{+0.4}_{-0.5}$\%&$-0.41^{+0.02}_{-0.02}$ & $ 1.35^{+0.04}_{-0.04}$& 0.65\\
\\\hline
%\\
\multicolumn{7}{l}{$^a$: parameter fixed}\\
\end{tabular}
\end{table*}

\section{Analysis and discussion}
\label{Sec:Dis}

The visibilities and closure phases obtained at the various epochs clearly reveal asymmetries that cannot be due to a single, spherical object, as shown by Fig.~\ref{Fig:vis2pionier}. 
We used 
the {\tt LITpro} software\footnote{{\tt LITpro} is available from \url{http://www.jmmc.fr/litpro\_page.htm}} \citepads{Litpro} to model the PIONIER data. 
Using the data of January 2015 (i.e. those with the most points), we tested several models, and it was clear that we need to have at least two components to best fit the data -- this is obvious from the clear signal in the closure phases. We show in Table~\ref{tab:models} the results for the best models we found: two  point-like sources, two discs, the same when adding a background source, as well as a case for a  point-like source, a disc and a background. It is clear that going from two points to two discs largely reduces the $\chi^2$ (from 3.6 to 1.8), while adding some background flux still decreases it to 1.5. Thus, 
for the continuum data observed by PIONIER, the best fit is given by two discs with a tiny background component. Fig.~\ref{Fig:chi2map} shows the reduced $\chi^2$ maps for the diameters of the two components as well as for the flux of the background. This shows the confidence intervals of these parameters. From these as well as from Table~\ref{tab:models}, it is clear that while the fluxes and the diameter of the primary are rather well constrained, that of the secondary is not, with errors of 20\% for the January 2015 dataset. We will come back to this below.
It is important to note that the relative positions of the two components is independent (within 3$\,\sigma$) of the model considered. This makes the derivation of the orbit (see below) very robust. 

We then applied the same model to all our epochs to measure the diameter of the two discs and the relative fluxes and positions (see Table~\ref{tab:models}). Rather noteworthy is that for the best model, the diameters of the two components are relatively stable, which leads us to think that they are possibly physical. The relative flux fraction of both the secondary and the background does appear to change with time and we will discuss this later.

We further used the {\tt CANDID} code\footnote{\url{https://github.com/amerand/CANDID}} \citepads{2015A&A...579A..68G}, a set of Python tools that was specifically made to search systematically for high-contrast companions. Although {\tt LITpro} and {\tt CANDID} are both based on a Levenberg-Marquardt minimisation, 
{\tt CANDID} performs a systematic exploration of the parameters while {\tt LITpro} only does one fit, and {\tt CANDID}
takes into account bandwidth smearing\footnote{But this latter aspect should not make much difference in the 
	  case of HR\,Car.}. 
 {\tt CANDID} provided the best fit when using a disc for the primary, an unresolved secondary and some background flux. The results for the various epochs are given in Table~\ref{tab:candid} 
-- it is clear that 
the derived values are in agreement with those found by {\tt LITpro}, typically within 1$\,\sigma$. For the March 2014 epoch, the data were not good enough for {\tt CANDID} to converge on a primary diameter and we fixed its value to the mean of the other epochs.

%\afterpage{
\begin{figure}[htbp!]
\includegraphics[width=9cm]{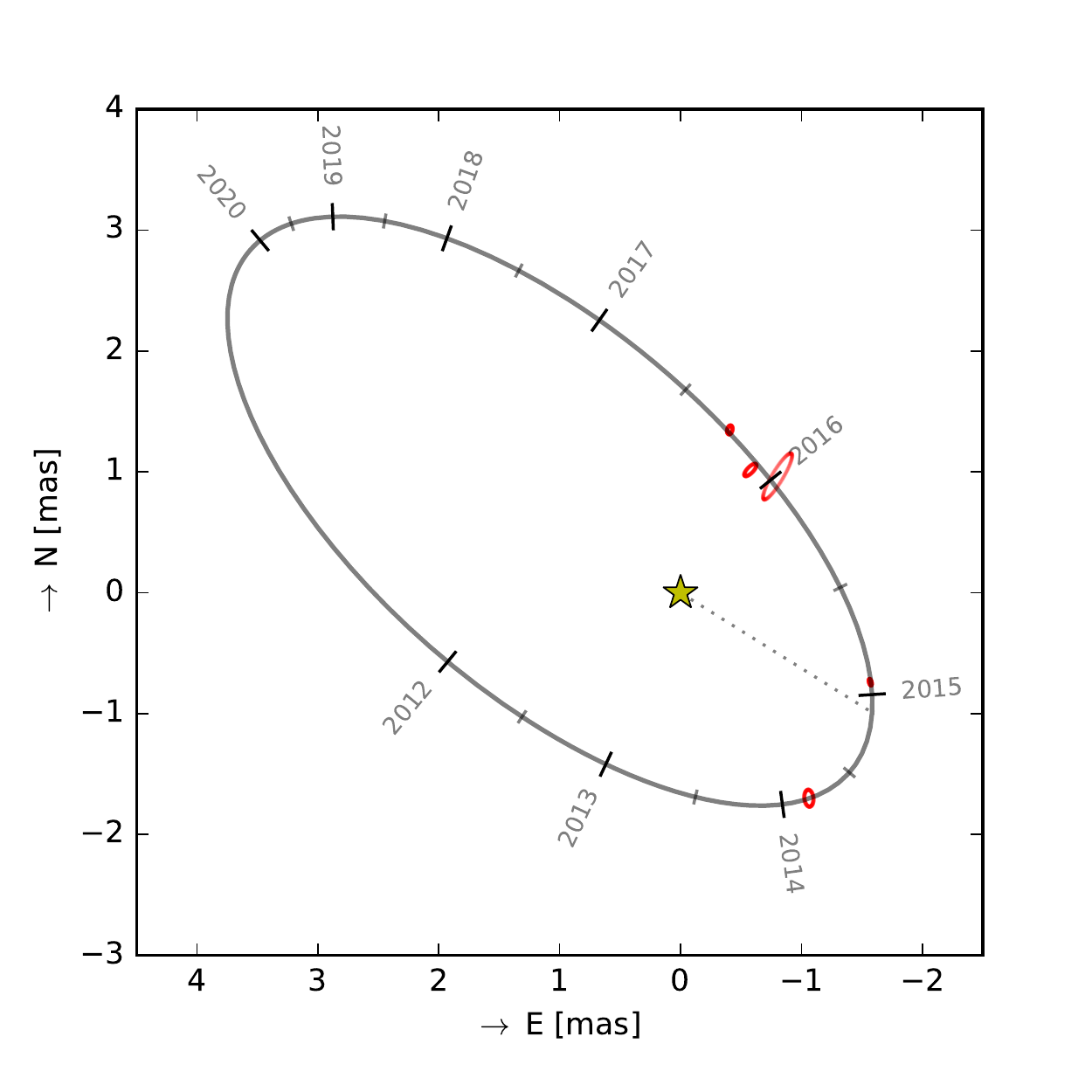}
\caption{\label{Fig:candid2} Best orbit obtained to fit all our data points. The parameters of this orbit are given in Tab.~\protect{\ref{tab:best orbit}}.}
\end{figure}
%}

%\afterpage{
 %\clearpage  
\begin{figure*}
\begin{center}
\includegraphics[width=18cm]{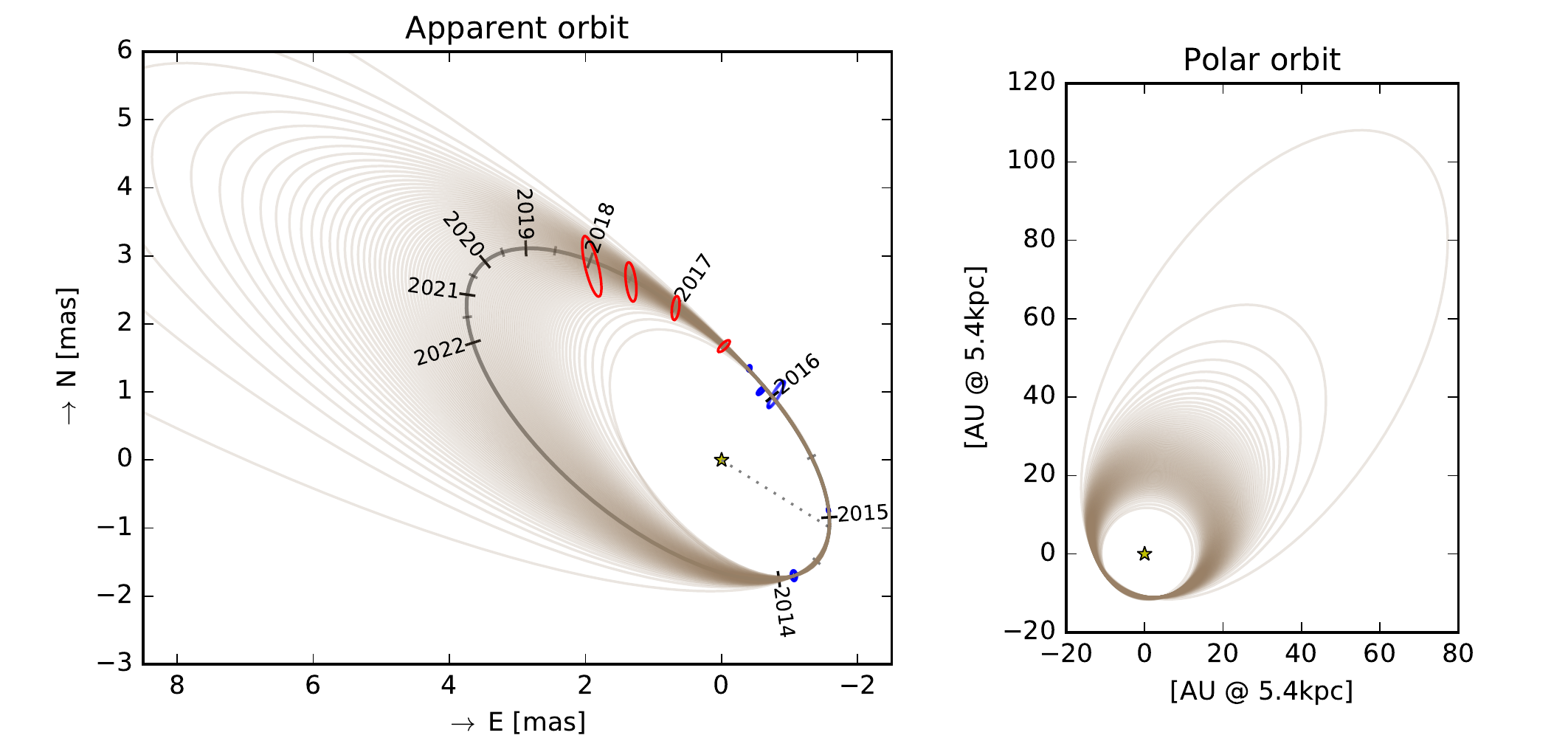}
\caption{\label{Fig:candid3} Family of orbits that fit all our PIONIER data points. The left panel shows the observed points and the projected orbits, as well as 1$\,\sigma$ ellipses allowing to distinguish between the various orbits. A large number of orbits still fit the data, with the longer orbits being the more eccentric. The right panel shows the de-projected orbits, in absolute scale, assuming a distance of 5.4 kpc. All orbits from Fig.~\ref{Fig:candid4} are shown here.}
\end{center}
\end{figure*}
 %\clearpage % prevent other material from being placed on this page
%} 

From this analysis, we are thus confident that the PIONIER data indicate:
\begin{itemize}
\item That there are two objects in the system;
\item That they are moving relative to each other in an apparent orbital motion;
\item That the relative flux fraction of the secondary in the $H$-band is about 15$\pm$1\% from March 2014 till February 2016, and thus that the flux ratio between the two components is 5.6$\pm$0.5, but that it dropped to 10\% in April 2016 (flux ration of 9);
\item That the primary has a diameter (in the $H$-band continuum) of 0.38$\pm$0.07 mas;
\item There is also some background flux, amounting to 2--3\% until January 2016, but increasing since February 2016 to reach about 5--6\%.
\end{itemize}

In addition, it is possible that the secondary is also resolved, although the disagreement between {\tt LITpro} and {\tt CANDID} on this indicates that the utmost caution is required in considering this possibility. If it is resolved, it has a diameter of 0.85$\pm$0.20 mas, i.e. the less luminous component is more than twice as big as its companion.
We will now discuss in turn what we can derive from these facts.

\begin{table}[htbp]
   \centering
    \caption{  \label{tab:best orbit} Parameters of the best orbit that fits the PIONIER points.}
   \begin{tabular}{@{} lc @{}} % Column formatting, @{} suppresses leading/trailing space
        \hline\hline
\\
$\Omega$ (deg.) & 46.9$\pm$0.6\\
Orbital period (days) &4557.5$\pm$21.0\\
$T_0$ (MJD) &   56990.6$\pm$16.0\\
semi-major axis (mas) &     3.324$\pm$0.026\\
eccentricity &    0.4$\pm$0.2\\
inclination (deg.) &    119.2$\pm$0.7\\
$\omega$ (deg.)& 201.9$\pm$2.1\\
  \hline
  \end{tabular}
\end{table}

\subsection{Orbit}

Although our derived positions that change with time clearly reveal that HR\,Car is a binary, we do not have yet enough data to fully constrain the orbit. Our data points, as obtained with {\tt CANDID}, are shown in Fig.~\ref{Fig:candid2}, together with the best orbit that fits our PIONIER data points -- i.e. the one with the smallest $\chi^2$. The corresponding parameters are given in Tab.~\ref{tab:best orbit}. This orbit has an orbital period of 12 years and an eccentricity of 0.4. At a distance of 5.4 kpc, the semi-major axis of this solution is 18 au. 

\afterpage{
\begin{figure*}[htbp]
\begin{center}
\begin{tabular}{cc}
\includegraphics[width=7.5cm]{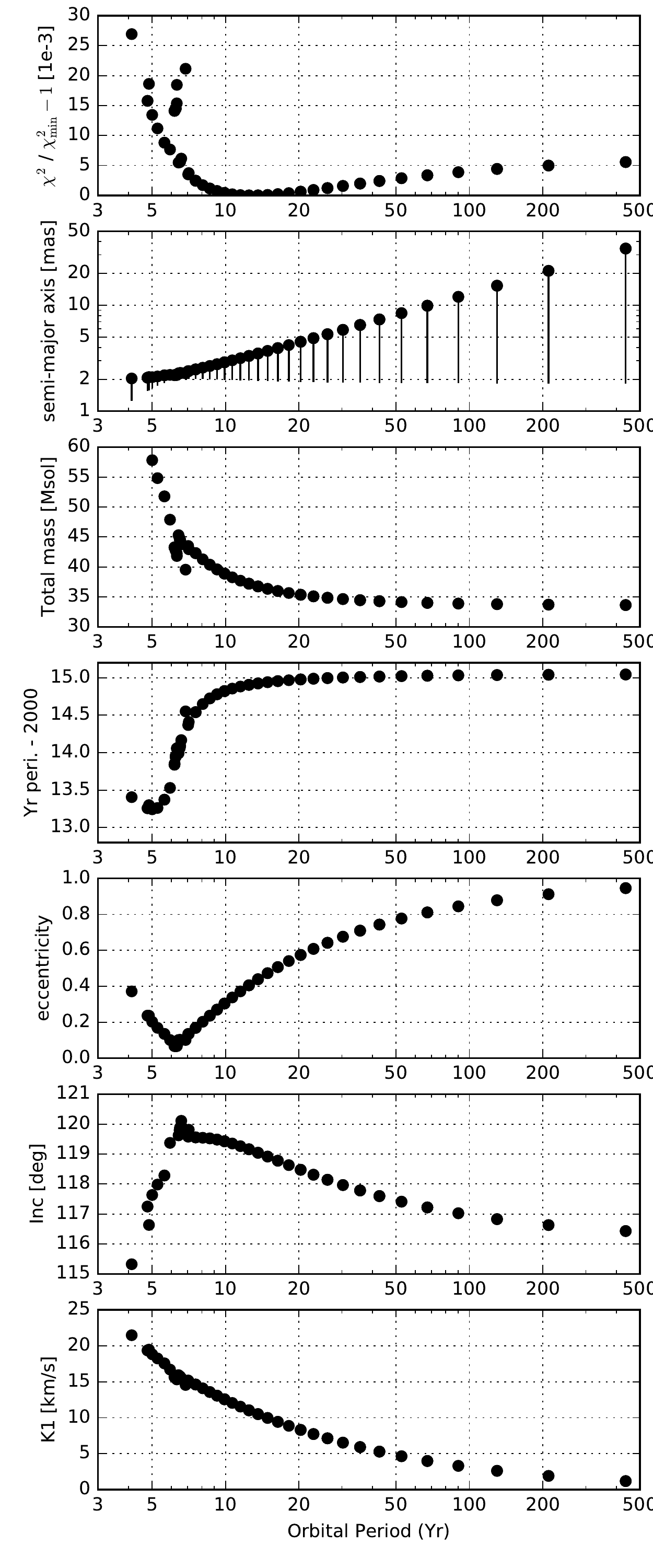} & 
\includegraphics[width=7.5cm]{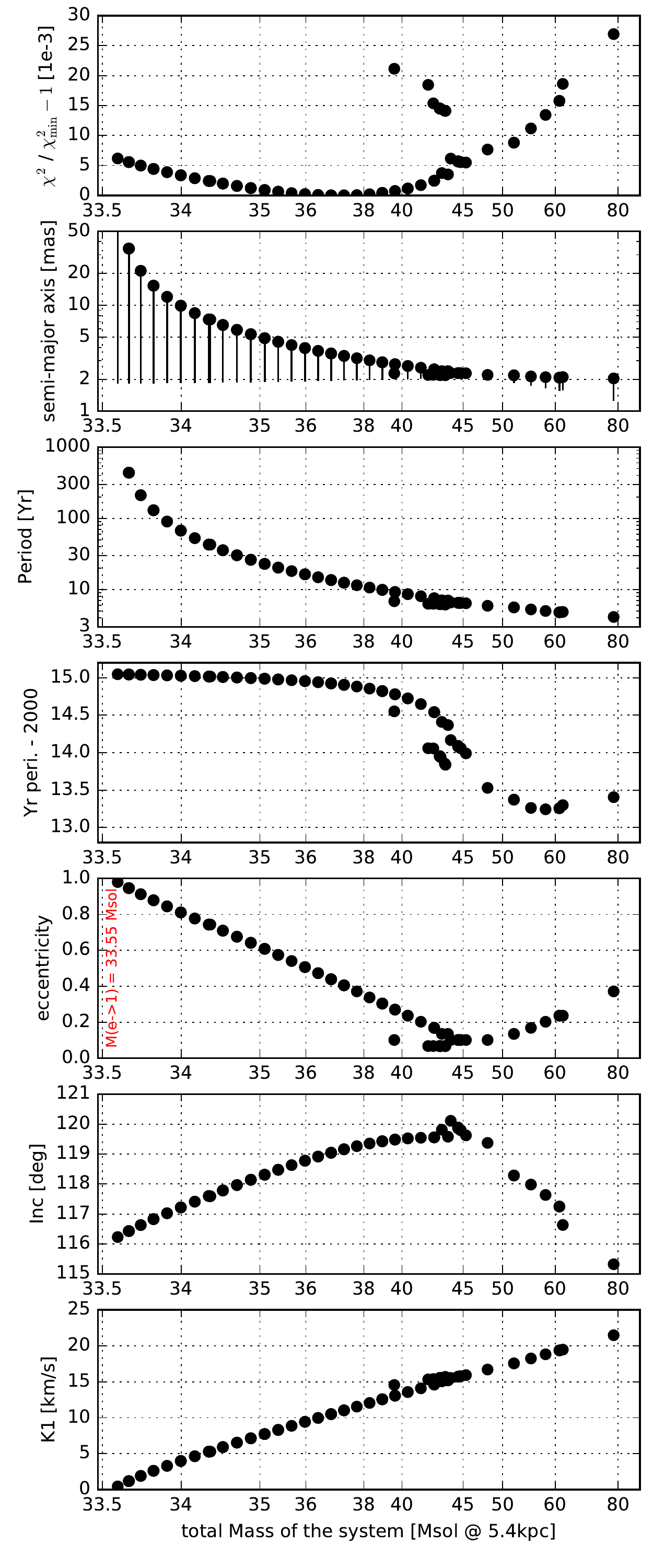}\\
\end{tabular}
\caption{\label{Fig:candid4}  Results from our solution grid-search to fit all the PIONIER data. The left panels show the parameters plotted as a function of the orbital period, while the right ones show them as a function of the total mass of the binary, assuming a distance of 5.4 kpc. The upper panels show the difference in ratio of the reduced $\chi^2$; the second the semi-major axis, with the line indicating the periastron distance; third is either the mass as a function of period or vice-versa; fourth panels are the time of periastron passage ($- 2000$); fifth, the eccentricity; the sixth show the inclination of the orbit, while the bottom panels show the radial velocities of the primary assuming a mass ratio of 2. Note the degeneracy in orbital solutions near zero eccentricity. These orbits have slightly worse $\chi^2$ than the others at same semi-major axis.
}
\end{center}
\end{figure*}
}

However, this is only one possible orbit, as clearly shown by Fig.~\ref{Fig:candid3}, which shows the result from a grid-search for all possible orbits, looking for the minimal $\chi^2$. From there, as well as from Fig.~\ref{Fig:candid4}, it appears that for now we cannot constrain the period and solutions between a few years up to several hundred years are possible. Of course, all parameters are not free and it is obvious that the long period orbits need to be very eccentric, a condition that is not required for shorter periods. In fact, given the arc we already have, we can be confident that the ({\it deprojected}) periastron is quite well defined around 2 mas, as is obvious from the second panels of Fig.~\ref{Fig:candid4} -- except for those with very short semi-major axis and high masses, but these are also those less likely in terms of $\chi^2$. This corresponds at  a nominal distance of 5.4 kpc to about 11 au. The longest periods thus correspond to the highest eccentricity, with the largest possible period\footnote{From an anthropocentric argument, the longest periods are also less favourable as they imply that we have been quite lucky to be observing the system now, when at its closest approach.} having an improbable eccentricity close to 1. For a small range of periods -- roughly 5 to 8 years -- the eccentricity is quite moderate and below 0.2. Those are not, however, the solutions which are the most likely in terms as $\chi^2$. 

The other parameters are also not independent, and they vary in a very narrow range. Thus the time of periastron passage is constrained between 2013.2 and 2015.05, the later being valid for most long-period solutions. Only for the shortest periods did the periastron happen earlier. The inclination of the orbit on the plane of the sky is also quite narrowly defined, between 115 and 120 degrees. 

Assuming a distance of 5.4 kpc, we can also translate our constraints as a function of the total mass of the binary systems and see that our solutions imply a total mass range between 33.6~M$_\odot$ and 80~M$_\odot$ (for the shortest orbits). The most favoured range (in terms of $\chi^2$) is the lowest one, i.e. roughly between 33.6 and 45~M$_\odot$. Assuming a mass of 25~M$_\odot$ for the primary LBV (see above), this would imply a companion mass between roughly 9 and 20~M$_\odot$, i.e. a mass ratio between 0.36 and 0.8. 

In Fig.~\ref{Fig:candid3}, we show the ellipses corresponding to 1$\,\sigma$ difference between the various orbits. If the size of these ellipses is much larger than the precision on our positions, then we should be able to distinguish between the different families. From the figure, it seems clear that by 2018, we should already be able to much better constrain the orbit and tell if HR\,Car is a short- or long-period binary.

Depending on the orbital period and eccentricity, one could also estimate the possible velocity change that the primary would undergo over its orbit. This of course depends crucially on the mass ratio in the system, and for illustration purpose we have here assumed a value of 0.5. As seen from the bottom panels of Fig.~\ref{Fig:candid4}, the resulting semi-amplitude of the primary is in the range between a few to 22 kms$^{-1}$ (for the shortest orbits). Given that HR\,Car is characterised by extremely variable lines over a S\, Dor cycle, as well as outflows of up to 150 kms$^{-1}$, such orbital motion would be hard to detect, explaining most likely why HR\,Car was not yet characterised as a spectroscopic binary.

Finally, it is important to realise that the binary companion we detected here has nothing to do with the possible B0 V companion inferred by \citetads{2000ApJ...539..851W} from radio data, as this companion is thought to be about 2 arcseconds away, i.e. a thousand times farther away than the one we detected.

\subsection{Size of the primary}

For the primary, we derive an angular  $H-$band diameter of the primary of 0.38$\pm$0.08 mas. At the distance of 5.4$\pm$0.4 kpc, this would translate in a radius of 220$\pm$60~R$_\odot$, which is nominally too large for the radius of the LBV\footnote{Given the bolometric magnitude and temperature quoted in the Introduction, we derive a radius of $90\pm25$~R$_\odot$, much smaller than what is found here. Nevertheless, the radius we derive is about 2.5 larger, which is compatible with the size of the wind continuum emission region.}. However, one should not forget that LBVs are generally surrounded by large envelope of ejected matter \citepads[e.g.][]{2015A&A...578A.108V}. It is in fact well known that HR\,Car is surrounded by a bipolar nebula whose lobes have a diameter of 0.65 pc, i.e. much larger than anything we measure here  \citepads{1997A&A...320..568W}. One can thus assume that we are here detecting an optically thick wind around the photosphere of the star.
Moreover, we are seeing the star during its current S Dor minimum -- its radius should be much larger during a maximum. We speculate that it is perhaps because we are in a minimum that we were able to detect such a faint companion -- it was with a flux ratio of 6 when we discovered it and is now with a flux ratio of 9, but during a maximum, the flux ratio would be in the range 30--40!

\subsection{Size and possible nature of the secondary}

We derive a rather constant angular diameter for the secondary companion of 0.85$\pm$0.20 mas, translating to a radius of 500$\pm$150~R$_\odot$, while at the same time, this object is about 6 to 9 times fainter in the $H-$band than the primary LBV star. This seems a priori difficult to reconcile, unless, perhaps, if one considers the possibility that the secondary is a red supergiant. 
Using the current $V\sim9$ magnitude of the combined HR\,Car system, and values of $V-H$ and bolometric corrections \citepads{2005ApJ...628..973L,2013ApJ...767....3D} of red supergiants, it is easy to show that a typical red supergiant with an effective temperature of 3,600--4,000 K would have the right flux and size to fit the data. Such a star would have a mass below 15~M$_\odot$, and would contribute to less than 1/10th of the total flux of the system in the $V$-band, explaining why it had escaped detection up to now. It remains to be seen if a binary system of a given age with a massive LBV and a less massive red supergiant can be reproduced by stellar evolutionary models, but this is beyond the scope of this paper.

Another possibility is that the secondary is out of equilibrium. The LBV was much larger around the year 2001 (around the maximum), so if the periastron happened close to this epoch\footnote{A possibility to consider if the orbital period is of the order of a few to 15 years.}, the secondary could have gone though the outer layers of the LBV, possibly leaving the secondary out of hydrostatic equilibrium. 

As the sum of the radii of both components we determine is about 0.6 mas, this means that even at periastron, the separation between them is more than thrice as much. The smallest Roche lobe radius for the secondary would be about 0.76 mas, which is much more than the radius we measure for the secondary. Hence, the secondary will never be filling its Roche lobe. 

One should, however, note that the flux of the secondary is varying with time -- it appears to be decreasing from a relative contribution of 16\% to 10\% within the two year window of our observations. This could mean that the primary is becoming brighter, but this not confirmed by the AAVSO light curve which shows that -- in the $V$-band -- the flux of HR\,Car has only varied by 30\% at most since January 2014. 
On the other hand, as shown earlier, the periastron passage took place most likely in late 2014 for a wide range of orbital periods -- so it is possible that the flux of the secondary varies due to some interaction with its companion or due to some geometrical effect, and that it is now weakening as the companion moves further away from the primary. If this were the case, one should not try to relate the flux we measure to a bolometric magnitude of the secondary.

\subsection{Background flux}

Our best fit clearly requires the presence of some background flux, at the level of a few percents. We are unable with the current data to determine the size of the region that contributes to this background flux, but it has to be larger than the separation between the two components. It is tempting to relate this background flux to what AMBER detected, i.e. the Br$\gamma$ emission region, although one should note that in the continuum, and at shorter wavelength, the wind extension must be much smaller than the line formation region.

Most interesting is the fact that the flux of the background has greatly increased since January 2016 -- in fact, it almost tripled in about three months! It will be interesting to follow the evolution of this background flux and in particular try to understand how it relates to the orbital parameters.

We would like to stress, however, that it is possible that the background flux we measure is an instrumental effect, as there has been some claims that PIONIER has possibly a bias in the visibilities for bright objects done with the HIGH gain. Until this can be fully discarded, we should beware of over-interpreting this background.

\subsection{Return to AMBER}
With the binary nature of HR\,Car confirmed and the system dimensions
constrained with PIONIER, revisiting the original AMBER data offers further
insights.  The first PIONIER observation was simultaneous to the OHANA survey data
and confirms that the phase signature in the Br$\gamma$ emission marks an
offset of the emission towards the secondary. Further, the OHANA observations
show a clearly elongated wind, larger in the direction perpendicular to the
line-of-sight between the components. Finally, the measured projected distance
between the components in March 2014, 2\,mas, is less than the extension of
the Br$\gamma$ emission region.

It is quite straightforward to reconcile that description with the images
evoked by SPH simulations of wind-wind interaction  (if the secondary star has a wind), as for instance shown by
\citetads{2013MNRAS.436.3820M} for $\eta$\,Car. The wind-wind interaction front
 may reduce the size of the emission region along the line-of-sight between the
components, and at the same time increases the local emissivity. As a
prediction for an eventual detection of the orbital RV curve, it follows that
during the 2014 observations, the secondary was closer to Earth than the
primary, so that the wind-wind interaction took place in the blue-shifted part
of the primary's dense wind.

Although the orbit is not yet fully constrained, a few things are already
clear. In the case of an eccentric system, with a rather long period of well
above a decade, the 2014 observations must have been around periastron,
within about a year, while the 2010 observations were not. The projected
periastron distance between both components is then the distance measured,
about 2\,mas. This would be one possibility to explain the absence of a phase
signature and the more extended wind NE-SW wind in 2010, namely that in 2010
the undisturbed configuration of the wind was observed, which is similar to
the one measured in 2014 along the NW-SE axis.

In the case of a circular orbit, which would have a shorter period of up to
about a decade, the time difference between the observations is such that the
wind-wind interaction front might just have rotated by about one quarter, so
that the 2010 $u,v$ position of panel e) is equivalent to the 2014 $u,v$
position\footnote{Or its conjugated point, as since there is no phase
signature, this is degenerate.} of panel d).

\section{Conclusions}
We have obtained interferometric observations of the LBV HR\,Car that clearly reveals its binary nature, and detected the orbital motion over a period of two years. It is still not possible to derive the orbital period which could be of the order of a few to several tens of years and the separation of the order of 10--270 au, but with the constraint that the largest orbit must also be the most eccentric, with a periastron distance most likely fixed around 2 mas, or 11 au. If the eccentricity is small and the orbit turns out to be of the order of 5 to 10 years, {\bf HR\,Car would be the second binary LBV presenting all the hallmarks and properties which make $\eta$~Car truly such a unique object}, but with components of much smaller masses. We should note, however, that no giant eruption has been ever recorded for HR\,Car, unlike the one of the 1840's of $\eta$~Car, and that estimates of the ejecta mass surrounding HR\,Car are more of the order of 1~M$_\odot$ \citepads{2000ApJ...539..851W}, much smaller than what is seen around $\eta$~Car.

Apart from highlighting the possible role of binarity in the formation and/or evolutions of LBVs, the fact that HR Car is a binary is essential as it will allow us to derive the masses of the stars, which will be very useful to  compare to stellar evolutionary models. For now, we constrain the most likely range of total masses to be 33.5--45~M$_\odot$. 

AMBER has shown that the emission line forming region of Br$\gamma$ is larger
than the minimum projected separation of the components of 2\,mas, measured by
PIONIER. Hence HR\,Car must undergo wind-wind interaction detectable in
Br$\gamma$, and probably as well in H$\alpha$.  Whether the interaction is
permanent, in case of a circular orbit, or phase dependent at periastron, in
case of an eccentric orbit, cannot be said with the current AMBER data, although the increase in background flux seen in the PIONIER data seem to favour the latter (if it proves to be non-instrumental). In either
case, however, the HR\,Car system is considerably simpler than its much better
known, nearby LBV-sibling, $\eta$\,Car, and probably much easier to constrain,
model, and ultimately understand. In particular in the case of an eccentric
orbit the next periastron would offer an excellent opportunity for a concerted
multi-wavelength, multi-technique campaign to provide constraints for
theoretical modelling.

Whatever interaction happens in the system of HR\,Car now, in the minimum
phase of its S\,Dor cycle, it must be very different when it is at maximum. In
the maximum of an S\,Dor cycle the primary, well separated from the secondary
now, even at the closest distance, will possibly become close or exceed its Roche-lobe radius, and
maybe even become large enough for the secondary to pass through the primary's
outer layers. In the recent past, two S\,Dor cycles have been observed for
HR\,Car, with maxima around 1991 and 1999 \citepads[see,
  e.g.,][]{2003IAUS..212..243S} or 2001 (as indicated by the AAVSO data). Once the orbit is better constrained, it will
be seen how these dates relate to the orbital parameters; whether one should
have expected strong interaction between the components, and in particular
what to expect in the next S\,Dor maximum phase.

We will continue to monitor HR\,Car with the PIONIER instrument to try to settle as soon as Nature allows us the orbital period of this interesting binary -- from the current modelling, in 2018 we should already be able to distinguish between the main families of solutions. This combined with a precise GAIA distance should also constrain the total mass of the system. We encourage spectroscopic monitoring of this target to try to derive the associated spectroscopic orbit, although we understand that this won't be a task for the faint-hearted.

\begin{acknowledgements}
It is a pleasure to thank Steve Ertel for taking the data in January 2015 in delegated Visitor Mode. The archival AMBER observation was obtained under ESO Prog. ID 085.D-0490.
\end{acknowledgements}

\end{document}